\documentclass[runningheads]{lncse}
\usepackage{multicol} 
\usepackage{makeidx}  
\usepackage{amssymb}
\usepackage{amsmath}
\usepackage{epsfig}
\newcommand{\rb}[1]{\raisebox{1.5ex}[-1.5ex]{#1}}
\begin{document}
\title{Density-Matrix Algorithm for Phonon Hilbert Space Reduction
in the Numerical Diago- nalization of Quantum Many-Body Systems}
\titlerunning{Density-Matrix Algorithm for Phonon Hilbert Space Reduction}

\author{Alexander Wei{\ss}e\inst{1}, Gerhard Wellein\inst{2}, 
\and Holger Fehske\inst{1}}
\authorrunning{Wei{\ss}e, Wellein and Fehske} 
\institute{Physikalisches Institut, Universit\"at Bayreuth, D-95440 Bayreuth
\and
Regionales Rechenzentrum Erlangen, Universit\"at Erlangen, D-91058 Erlangen}

\maketitle              
\index{Wei{\ss}e@Alexander}
\index{Wellein@Gerhard}
\index{Fehske@Holger}
\def\ho{\omega_0}
\def\cH{{\cal{H}}}
\def\ep{\varepsilon_p}
\def\eps{\tilde{\varepsilon}_p}
\begin{abstract}
Combining density-matrix and Lanczos algorithms we propose 
a new optimized phonon approach for finite-cluster 
diagonalizations of interacting electron-phonon systems. 
To illustrate the efficiency and reliability of our method,
we investigate the problem of bipolaron band formation 
in the extended Holstein Hubbard model.
\end{abstract}
\section{Introduction}
Considerable work is currently focused on the experimental 
and theoretical study of strongly coupled electron-phonon (EP)  
systems, triggered by the recognition that the EP interaction 
plays an important role in understanding the physics 
of novel materials such as colossal magneto-resistance 
manganites~\cite{JTMFRC94} or the very recently discovered 
superconducting magnesium diboride~\cite{NNMZA01}.
From a theoretical point of view the challenge is to 
describe the partly exotic properties of these materials
in terms of simplified microscopic models which take into account
the complex interplay of charge, spin and lattice degrees of freedom.  

As a generic model for systems with competing 
electron-electron and electron-phonon interactions 
the extended Holstein Hubbard model (EHHM), 
\begin{equation}
\cH\!=\!-t\!\sum_{\langle i,j\rangle;\sigma }\! c_{i\sigma}^{\dagger} c_{j\sigma}^{}
+U\!\sum_{i}\!n_{i\uparrow}n_{i\downarrow}
-\sum_{i,l;\sigma} f_l(i)n_{i\sigma} x_0 ( b_l^{\dagger}  + b_l^{})+\ho\sum_i  ( b_i^{\dagger} b_i^{}+\mbox{\small $\frac{1}{2}$}), 
\label{ED1}
\end{equation}
is usually considered~\cite{AK99,FLW00,BT01}, 
where $c_{i\sigma}^{[\dagger]}$ and 
$ b_i^{[\dagger]}$ annihilates [creates] a spin-$\sigma$ electron and 
a phonon at Wannier site $i$, respectively, and  
$n_{i\sigma}^{}=c_{i\sigma}^{\dagger}c_{i\sigma}^{}$.
The Hamiltonian~(\ref{ED1}) consists of a kinetic term describing the
electronic motion on a discrete lattice (transfer amplitudes $t$), 
an extremely screened (on-site) Coulomb repulsion (Hubbard parameter $U$),  
and a ``density-displacement'' type non-screened EP coupling 
($\propto \kappa x_0$)
\begin{equation}
f_l(i)=\frac{\kappa}{(|l-i|^2+1)^{3/2}}\;,\quad x_0=\sqrt{1/2M\ho}\;,
\quad \kappa x_0=\sqrt{\ep\ho}\,.
\label{krako} 
\end{equation}
Here $\ho$ is the bare phonon frequency of dispersionsless 
optical phonons, being polarized in the direction perpendicular 
to the chain (1D case). Defining the polaron binding energy as
$\eps=(x_0^2/\ho) \sum_lf^2_l(0)=1.27\ep$, 
the famous Holstein Hubbard model (HHM) results by setting
\begin{equation}
f_l(i)=\kappa \delta_{i,l}\;,\quad \eps \to \ep\,,
\label{HHM}
\end{equation}
i.e., with respect to the EP coupling term the EHHM
represents an extension of the Fr\"ohlich model~\cite{Fr54} 
to a discrete ionic lattice or of the Holstein model~\cite{Ho59a} 
including longer ranged EP interactions.

Adapting the EHHM to real physical situations one is frequently 
faced with the difficulty that the energy scales of 
electrons~($t,\ U$), phonons~($\ho$) and their 
interaction~($\eps$) are of the same order of magnitude, causing 
analytic methods, and especially adiabatic techniques, to fail 
in most of these cases. Thus, at present, the most reliable 
results came from powerful numerical calculations,  
such as finite-cluster exact  diagonalizations 
(ED)~\cite{RT92,AKR94,WF97} or  
(Quantum) Monte Carlo simulations~\cite{RL82,BVL95}, which are
usually performed on supercomputers. 
But even for these numerical approaches strong EP interactions 
are a demanding task, since they require 
some cut-off in the phonon Hilbert space. Starting with the work 
of White~\cite{Wh93} in 1993, during the last years a class of 
algorithms became very popular, which based on the
use of a so-called density matrix for the reduction of large Hilbert
spaces to manageable dimensions.

In the present paper, we will demonstrate that finite-cluster
diagonalization methods also benefit substantially from these ideas.
Along this line, in the next section we introduce an optimized phonon 
approach for the ED of electron-phonon problems. To exemplify this 
technique, we analyze the formation of bipolarons in Sec.~III.
Our main results are summarized in Sec.~IV.
\section{Optimized phonon approach}
Let us first resume the connection between 
density matrices and optimized basis states. Starting with
an arbitrary normalized quantum state 
\begin{equation}
  |\psi\rangle = \sum_{r=0}^{D_r-1} \sum_{\nu=0}^{D_\nu-1} \gamma_{\nu r} |\nu\rangle|r\rangle
\end{equation}
expressed in terms of the basis $\{|\nu\rangle|r\rangle\}$ of
the direct 
product space $H=H_\nu\otimes H_r$, we wish to reduce the dimension $D_\nu$ 
of the space $H_\nu$ by introducing a new basis, 
\begin{equation}
  |\tilde\nu\rangle = \sum_{\nu=0}^{D_\nu-1} 
  \alpha_{\tilde\nu \nu}^{} |\nu\rangle\,,
\end{equation}
with $\tilde\nu=0\ldots (D_{\tilde\nu}-1)$ and $D_{\tilde\nu}<D_\nu$.
The projection of $|\psi\rangle$ onto the corresponding subspace 
$\tilde H = H_{\tilde\nu}\otimes H_r \subset H$ is given by
\begin{eqnarray}
  |\tilde\psi\rangle & = & 
  \sum_{r=0}^{D_r-1} \sum_{\tilde\nu=0}^{D_{\tilde\nu}-1} 
  \sum_{\nu'=0}^{D_\nu-1} 
  \alpha_{\tilde\nu \nu'}^{*} \gamma_{\nu' r}^{} 
  |\tilde\nu\rangle|r\rangle\nonumber{}\\
  & = & \sum_{r=0}^{D_r-1} 
  \sum_{\tilde\nu=0}^{D_{\tilde\nu}-1} \sum_{\nu,\nu'=0}^{D_\nu-1} 
  \alpha_{\tilde\nu \nu}^{} \alpha_{\tilde\nu \nu'}^{*} \gamma_{\nu' r}^{}
  |\nu\rangle|r\rangle\,.
\end{eqnarray}
We call $\{|\tilde\nu\rangle\}$ an optimized basis, if $|\tilde\psi\rangle$ 
is as close as possible to the original state $|\psi\rangle$. 
Therefore we minimize $\| |\psi\rangle - |\tilde\psi\rangle \|^2$ with 
respect to the elements $\alpha_{\tilde\nu \nu}$ of the transformation
matrix $\boldsymbol{\alpha}$ under the orthogonality condition 
$\langle \tilde\nu'|\tilde\nu\rangle = 
\sum_{\nu=0}^{D_\nu-1} \alpha_{\tilde\nu' \nu}^{*} \alpha_{\tilde\nu \nu}^{} = 
\delta_{\tilde\nu' \tilde\nu}$. Applying the latter condition, we find
\begin{eqnarray}
  \| |\psi\rangle - |\tilde\psi\rangle \|^2
  & = & \langle\psi|\psi\rangle - \langle\tilde\psi|\psi\rangle 
  - \langle\psi|\tilde\psi\rangle + \langle\tilde\psi|\tilde\psi\rangle
  \nonumber{}\\
  & = & 1 - \left[
    \sum_{r=0}^{D_{r}-1} \sum_{\tilde\nu=0}^{D_{\tilde\nu}-1} 
    \sum_{\nu,\nu'=0}^{D_{\nu}-1} \gamma_{\nu r}^{*} \alpha_{\tilde\nu \nu}^{} 
    \alpha_{\tilde\nu \nu'}^{*} \gamma_{\nu' r}^{} + \textrm{H.c.}
  \right]\nonumber\\
  & & +\ \sum_{r=0}^{D_{r}-1} \sum_{\tilde\nu,\tilde\nu'=0}^{D_{\tilde\nu}-1} 
  \sum_{\nu,\nu',\nu''=0}^{D_{\nu}-1} \gamma_{\nu'' r}^{*}
  \alpha_{\tilde\nu' \nu''}^{} \alpha_{\tilde\nu' \nu}^{*} 
  \alpha_{\tilde\nu \nu}^{} \alpha_{\tilde\nu \nu'}^{*} 
  \gamma_{\nu' r}^{}\nonumber\\
  & = & 1 - \sum_{r=0}^{D_{r}-1} \sum_{\tilde\nu=0}^{D_{\tilde\nu}-1} 
  \sum_{\nu,\nu'=0}^{D_{\nu}-1} \alpha_{\tilde\nu \nu}^{} \gamma_{\nu r}^{*} 
  \gamma_{\nu' r}^{} \alpha_{\tilde\nu \nu'}^{*} \nonumber\\
  & = & 1 - \textrm{Tr}(\boldsymbol{\alpha} \boldsymbol{\rho} 
  \boldsymbol{\alpha}^{\dagger})\,,
\end{eqnarray}
where $\boldsymbol{\rho} = 
\sum_{r=0}^{D_r-1} \gamma_{\nu r}^{*} \gamma_{\nu' r}$ is
called the {\em density matrix} of the state $|\psi\rangle$ with
respect to $\{|\nu\rangle\}$. 
We observe immediately that the states $\{|\tilde\nu\rangle\}$ are 
optimal if the rows of $\boldsymbol{\alpha}$ are eigenvectors of 
$\boldsymbol{\rho}$ corresponding to its $D_{\tilde\nu}$ largest 
eigenvalues $w_{\tilde\nu}$.

Following Zhang et al.~\cite{ZJW98}, we now apply these features to construct 
an optimized phonon basis for the eigenstates of an interacting 
electron/spin-phonon system. Consider a system composed of $N$ sites, 
each contributing a phonon degree of freedom 
$|\nu_i\rangle,\ \nu_i=0\ldots\infty$, and some other 
(spin or electronic) states $|r_i\rangle$. Hence, the Hilbert space
of the model under consideration is spanned by the basis
$\{\bigotimes_{i = 0}^{N-1} |\nu_i\rangle|r_i\rangle\}$. Of course,
to numerically diagonalize an Hamiltonian operating on this space, we
need to restrict ourselves to a finite-dimensional subspace. To
calculate, for instance, the lowest eigenstates of 
the HHM~(\ref{ED1})-(\ref{HHM}) 
we could limit the phonon space spanned by
$|\nu_i\rangle = (\nu_i!)^{-1/2} (b_{i}^{\dagger})^{\nu_i}|0\rangle$ by
allowing only the states $\nu_i<D_i$. Most simply we can choose
$D_i = M\ \forall\ i$ yielding $D_{\rm ph}=M^N$ for the dimension of the
total phonon space. However, if we think of the states 
$\{\bigotimes_{i = 0}^{N-1} |\nu_i\rangle\}$ as eigenstates of the 
Hamiltonian ${\cal H}_{\rm ph}=\omega\sum_{i=0}^{N-1} b_i^{\dagger} b_i^{}$, 
it is more suitable for most problems to choose an energy cut-off instead.
Thus we used the condition $\sum_{i=0}^{N-1} \nu_i < M$, leading
to $D_{\rm ph} = \binom{N+M-1}{N}$, for most of our previous numerical 
work (see e.g. Ref.~\cite{BWF98}).
For weakly interacting systems already a small number $M$ of phonon 
states is sufficient to reach very good convergence for ground states
and low-lying excitations. However, with increasing coupling strength most
systems require a large number of the above 'bare' phonons, thus
exceeding capacities of even large supercomputers. In some cases
one can avoid these problems by choosing an appropriate unitary
transformation of the Hamiltonian (e.g. by using center-of-mass coordinates), 
but in general it is desirable to find an optimized basis automatically.

\begin{figure}[!htb]
  \begin{center}
    \epsfig{file= 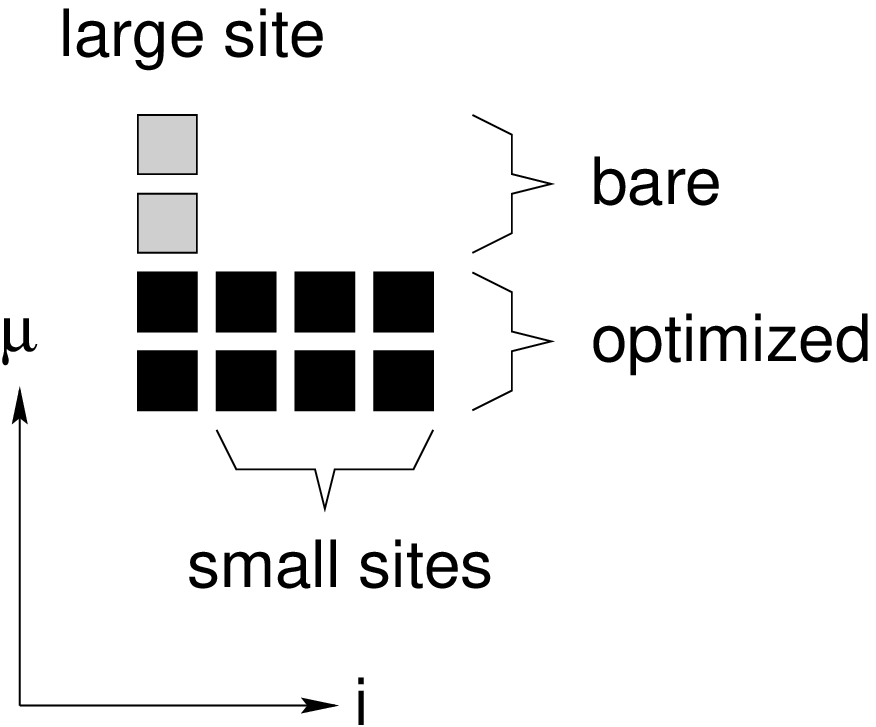, width=0.4\linewidth}
    \hspace{0.1\linewidth}
    \epsfig{file= 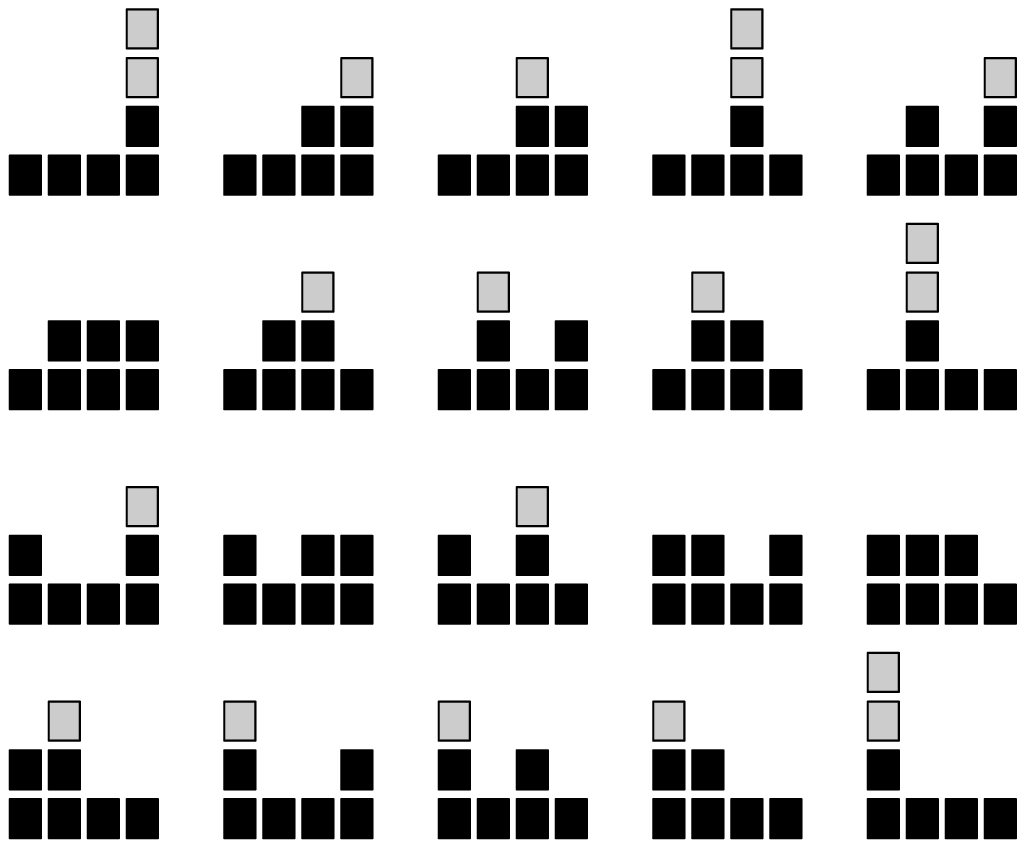, width=0.4\linewidth}
  \end{center}
  \caption{Structure of the phonon basis in terms of the highest
    accessible $\mu_i$. Left: as proposed by Zhang et~al.~\cite{ZJW98}; 
    Right: used within this work.}\label{figbase}
\end{figure}

Within the current density-matrix algorithm~\cite{ZJW98} for the construction 
of an optimal phonon basis the phonon subsystem is considered as a product 
of one 'large' ($i=0$) and a number of 'small' sites ($i>0$). 
Each site except the large one uses the same optimized basis 
$\{|\mu_{i>0}\rangle\} = \{|\tilde\nu\rangle\}$ 
with $\tilde\nu=0\ldots (m-1)$, while the basis of the large site consists 
of the states $\{|\tilde\nu\rangle\}$ plus some bare states 
$\{|\nu\rangle\}$, $\{|\mu_0\rangle\} = 
\textrm{ON}(\{|\tilde\nu\rangle\}\cup\{|\nu\rangle\})$, where 
$\textrm{ON}(\ldots)$ denotes orthonormalization (see Figure~\ref{figbase}). 
After a first initialization the optimized states are improved iteratively 
through the following steps
\begin{enumerate}
\item[(1)] calculating the requested eigenstate $|\psi\rangle$ of the 
  Hamiltonian ${\cal H}$ in terms of the actual basis,
\item[(2)] replacing $\{|\tilde\nu\rangle\}$ with the most important (i.e.,
largest eigenvalues $w_{\tilde\nu}$) eigenstates of the density matrix 
$\boldsymbol{\rho}$, calculated with respect to $|\psi\rangle$ and 
$\{|\mu_0\rangle\}$,
\item[(3)] changing the additional states $\{|\nu\rangle\}$ in the set 
$\{|\mu_0\rangle\}$,
\item[(4)] orthonormalizing the set $\{|\mu_0\rangle\}$, and returning to step (1).
\end{enumerate}
A simple way to proceed in step (3) is to sweep the bare states 
$\{|\nu\rangle\}$ through a sufficiently large part of the infinite 
dimensional phonon Hilbert space. One can think of the algorithm as 
'feeding' the optimized states with bare phonons, thus allowing the 
optimized states to become increasingly perfect linear combinations of 
bare phonon states. Of course the whole procedure converges only for 
eigenstates of ${\cal H}$ at the lower edge of the spectrum, since usually 
the spectrum of a Hamiltonian involving phonons has no upper bound. 
The applicability of the algorithm was demonstrated in Ref.~\cite{ZJW98} with 
the Holstein model (i.e., $U=0$ in Eq.~(\ref{ED1})) as an example.

When we implemented the above algorithm together with a Lanczos ED method 
for our systems of interest, we found two objections against the above 
choice of an optimized basis: 
(i) the basis is not symmetric under the symmetry operations of the 
Hamiltonian (e.g. translations), and 
(ii) the phonon Hilbert space  is still large ($D_{\rm ph} = M\,m^{N-1}$, 
where $M$ is the dimension at the large site), since we usually need more 
than one optimized state per site. 

\begin{figure}[!htb]
  \begin{center}
    \epsfig{file= 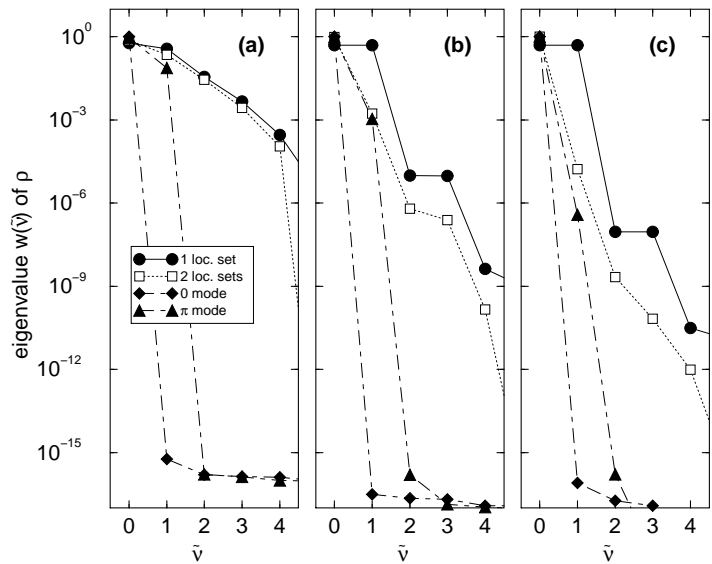, width=0.8\linewidth}
  \end{center}
  \caption{Eigenvalues $w_{\tilde\nu}$ of $\boldsymbol{\rho}$ 
    calculated with the ground state of the Holstein model 
    for $g=5$ and frequencies: (a) $\omega_0=0.1\,t$, (b) $\omega_0=t$, 
    and (c) $\omega_0=10\,t$. Filled circles: one local set; open squares: 
    two local sets for each fermion state; filled diamonds and triangles: 
    momentum dependent sets.}\label{figevals}
\end{figure}

The first problem is solved by including all those states into the 
phonon basis that can be created by symmetry operations (see 
Figure~\ref{figbase}, right panel), and by calculating the density 
matrix in a symmetric way, i.e., by adding the density matrices generated 
with respect to every site, not just site $i=0$.
Concerning the second problem we note that the eigenvalues $w_{\tilde\nu}$ 
of the density matrix $\boldsymbol{\rho}$ decrease approximately 
exponentially, see Figure~\ref{figevals}. If we interpret 
$w_{\tilde\nu}\sim \exp(-a\tilde\nu)$ as the probability of the system to 
occupy the corresponding optimized state $|\tilde\nu\rangle$, we immediately 
find that the probability for the complete phonon basis state 
$\bigotimes_{i = 0}^{N-1} |\tilde\nu_i\rangle$ is proportional to
$\exp(-a\sum_{i = 0}^{N-1} \tilde\nu_i)$. This is reminiscent of the
energy cut-off discussed above, and we therefore propose the following
choice of phonon basis states at each site,
\begin{eqnarray}
  \forall\ i:\ \{|\mu_i\rangle\} & = & \textrm{ON}(\{|\mu\rangle\})\\ 
  |\mu\rangle & = & \begin{cases}
    \textrm{opt. state }|\tilde\nu\rangle, & \ 0\le \mu < m \\
    \textrm{bare state }|\nu\rangle, & \ m\le \mu < M
  \end{cases}\,,
\end{eqnarray}
and for the complete phonon basis 
$\left\{\bigotimes_{\Sigma_i \mu_i < M} |\mu_i\rangle\right\}$,
yielding $D_{\rm ph} = \binom{N+M-1}{N}$.

\begin{figure}[!htb]
  \begin{center}
    \epsfig{file= 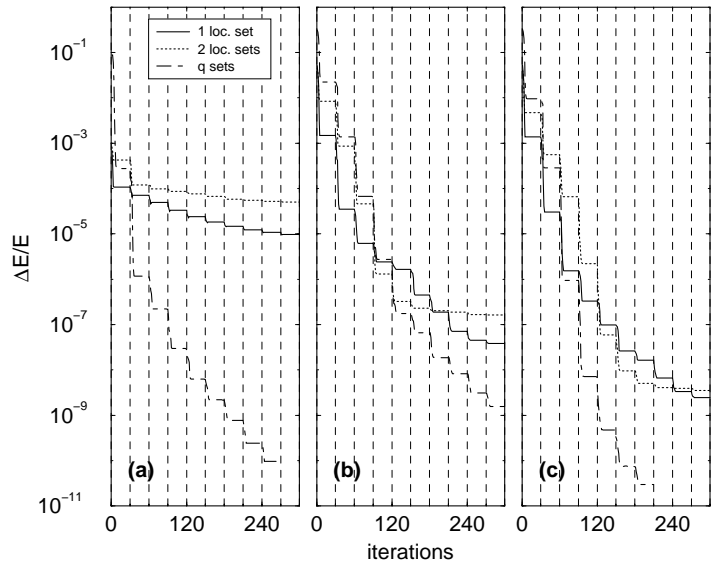, width=0.8\linewidth}
  \end{center}
  \caption{Convergence of the ground-state energy for the Holstein model, 
    Eq.~(\ref{Hsf}), with electron-phonon coupling $g=5$ and different 
    phonon frequencies: 
    (a) $\omega_0=0.1\,t$, (b) $\omega_0=t$, and (c) $\omega_0=10\,t$.
    Solid lines: one local set, dotted lines: two local sets for each
    fermion state, dot-dashed lines: momentum dependent sets.}\label{figconv}
\end{figure}

To discuss the nature of the obtained optimized states, the 
convergence of the algorithm, and some variants in more detail, 
let us consider a special case of the HHM, Eq.~(\ref{HHM}), namely 
the Holstein model of spinless fermions in one dimension,
\begin{equation}
  {\cal H} = 
  -t \sum_{i} \left[c_{i}^{\dagger} c_{i+1}^{} + \textrm{H.c.} \right]
  + g \omega_0 \sum_{i} (b_i^{\dagger} + b_i^{})(n_{i}-\tfrac{1}{2}) 
  + \omega_0 \sum_i b_i^{\dagger} b_i^{}\,.
  \label{Hsf}
\end{equation}
The optimized phonon approach comes into play in the case of strong 
electron-phonon coupling $g$. Then the systems develops lattice distortions
which accompany the itinerant fermions. These finite elongations
need to be expressed in terms of Harmonic oscillator states 
$|\nu_i\rangle = (\nu_i!)^{-1/2} (b_{i}^{\dagger})^{\nu_i}|0\rangle$,
that are centered around the equilibrium position. Hence, for the
numerical diagonalization either a large number of these 'bare' states 
or some other states embodying a finite distortion are required. 
By 'sweeping' through the large space of 'bare' phonons, the above 
optimization procedure automatically creates a small number of basis 
states $|\tilde\nu\rangle$, which are sufficient for a good approximation
of eigenstates of ${\cal H}(t,g,\omega_0)$.

In Figure~\ref{figconv} the convergence of the ground state energy of 
a two-site system at half-filling is shown for an increasing number of 
iterations (solid lines). We compare the results of an ordinary 
diagonalization using up to $M=80$ bare phonons per site (i.e.,
$D_{\rm ph} = \binom{N+M-1}{N} = 3240$) and of the optimized approach.
In the latter case the phonon basis consists of 6 optimized and 4 bare 
states, i.e., $M=10$ and $D_{\rm ph} = 55$. 
Each optimized state is chosen  to be a linear combination of the 
first 120 bare states. Initially we set $|\tilde\nu\rangle = |\nu\rangle$ 
and then sweep the 4 bare states through the states $|\nu\rangle$ with 
$\nu = 0\ldots 119$.
Vertical dashed lines denote the end of each sweep. The plateau structure
is due to the fact that states of high $\nu$ are less important
for the optimized basis $|\tilde\nu\rangle$. Note that every iteration
involves the calculation and diagonalization of the density matrix, 
an update of the operators $b_i^{(\dagger)}$, which need to be transformed 
to the current basis, and, most expensive, a Lanczos iteration to obtain
a new approximation for the requested eigenstate of ${\cal H}$.
It is therefore recommended to use the optimized states obtained for a 
small cluster as the initial basis for a larger cluster, and to 
restart the Lanczos procedure with the previous eigenvector (although
it is expressed in a slightly different basis).
 
The figure also includes data for two variants of the algorithm, 
namely the construction of {\em two} sets of optimized states, 
one for each local fermion state, or a transfer of the calculation
into {\em momentum space} and the use of different optimized 
states for each momentum $q$. 
The advantages of these ideas become clear looking at Figure~\ref{figevals}. 
Here the eigenvalues $w_{\tilde\nu}$ of the density matrix $\boldsymbol{\rho}$
are given for different phonon frequencies $\omega_0$ and coupling $g=5$.
In the anti-adiabatic regime of high frequencies we observe pairs
of equally important eigenstates of $\boldsymbol{\rho}$ [filled circles
in panels (b) and (c)]. This indicates that the lattice follows the fermion 
immediately  (small polaron). It is locally distorted to one side or the other, 
if the site is occupied by an electron or not. Hence, only half of the 
optimized states couple to one of the local fermion states, and it appears 
reasonable to use different basis sets for the two situations. 
This results in step free exponential decrease of the eigenvalues of
the density matrices for both basis sets (open squares in 
Figure~\ref{figevals}), which allows for a smaller cut-off $M$ in the 
total phonon space of the cluster.

However, at small frequencies the lattice is slow compared to 
the fermions, and distortions are long ranged. Obviously, there is 
no gain using different local basis sets, the convergence as well as
the decay of the eigenvalues $w_{\tilde\nu}$ of the density matrix 
is slow (see panel (a) in Figures~\ref{figconv} and~\ref{figevals}). 
The performance of each {\em local} optimization seems to be poor 
and switching to momentum space is recommended. After a shift of
operators, $b_i^{(\dagger)} \rightarrow b_i^{(\dagger)} + \tfrac{g}{2}$, and Fourier transform 
the Hamiltonian~${\cal H}$, Eq.~(\ref{Hsf}), reads
\begin{eqnarray}
  {\cal H}
  & = & -t \sum_{i} \left[c_{i}^{\dagger} c_{i+1}^{} + \textrm{H.c.} \right]
  + g \omega_0 \sum_{i} (b_i^{\dagger} + b_i^{}) n_{i} 
  + \omega_0 \sum_i b_i^{\dagger} b_i^{} + E_s\\
  & = & -2 t \sum_{k} \cos\left[\tfrac{2\pi}{N} k\right]\, n_k 
  + \frac{g \omega_0}{\sqrt{N}} \sum_{k,q} (b_{-k}^{\dagger} + b_{k}^{}) 
  c_{q+k}^{\dagger} c_{q}^{} 
  + \omega_0 \sum_{k} b_k^{\dagger} b_k^{} + E_s\,,\nonumber{}
\end{eqnarray}
with $E_s = g^2 \omega_0 \left(\sum_{k} n_{k} - \tfrac{N}{4}\right)$. 
Using the same cut-off in the total phonon basis as described before, 
we now optimize each $q$-mode separately. The results turn out to be 
excellent for {\em all} phonon frequencies. The ground-state energy 
converges quickly (dot-dashed lines in Figure~\ref{figconv}) and only
a few optimized states are required, as can be seen from the 
rapidly decreasing eigenvalues of the density matrix (filled 
diamonds and triangles in Figure~\ref{figevals}).

\begin{figure}[!htb]
  \begin{center}
    \epsfig{file= 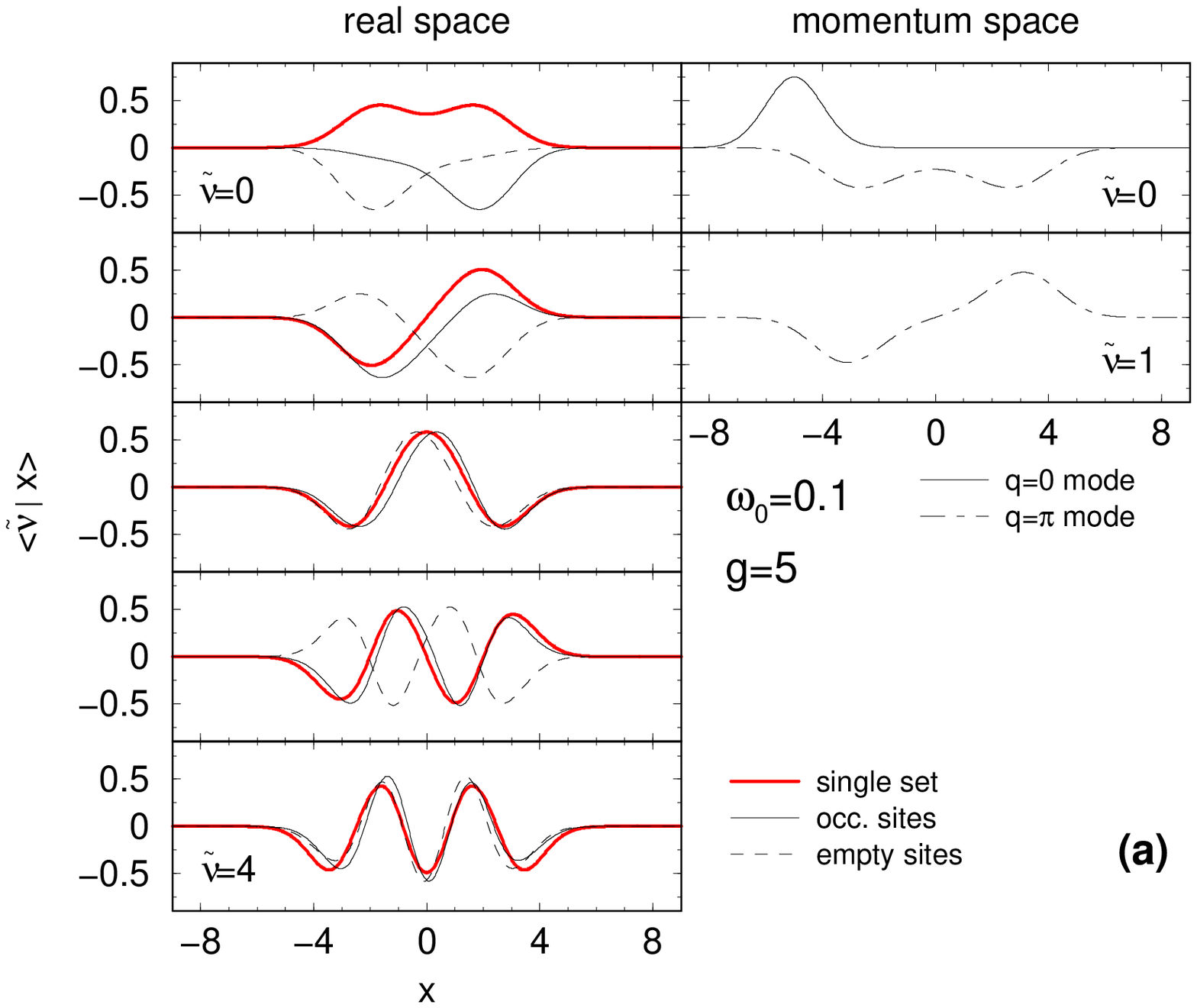, width=0.8\linewidth}
    \epsfig{file= 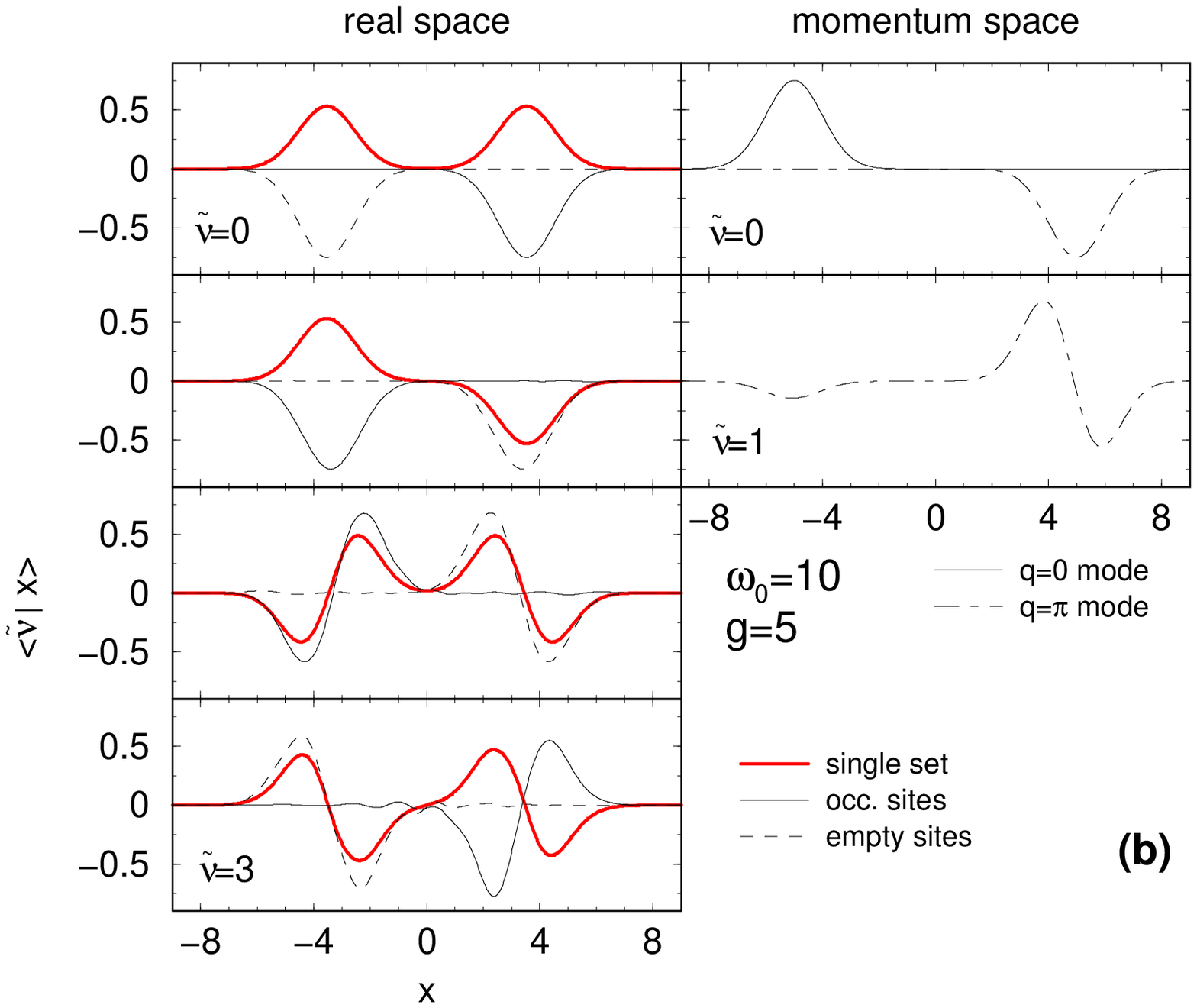, width=0.8\linewidth}
  \end{center}
  \caption{Optimized phonon states in elongation space for (a)
    $\omega_0 = 0.1\,t$, and (b) $\omega_0 = 10\,t$, and different
    choices of the optimized basis (left panels: real space; 
    right panels: momentum space).}\label{figwave}
\end{figure}

For illustration, in Figure~\ref{figwave} we give optimal wave
functions obtained for the frequencies $\omega_0 = 0.1\,t$ and $10\,t$
with $\tilde\nu$ increasing from top to bottom. In the left panels 
bold lines denote the wave functions of a single local set, 
whereas solid and dashed lines mark the two sets of functions that 
depend on the fermion occupation number. 
The right panels show optimal wave functions in momentum space.
In all cases $x$ denotes the normalized elongation 
$\tfrac{1}{\sqrt{2}}(b^{\dagger} + b)$, i.e., the expansion of the 
optimized states 
$|\tilde\nu\rangle = \sum_{\nu} \alpha_{\tilde\nu \nu}^{} |\nu\rangle$
is plotted using Harmonic oscillator eigenfunctions in elongation space,
\begin{equation}
  |\nu\rangle = \frac{e^{-x^2/2}}{\sqrt{2^\nu \nu! \sqrt{\pi}}} H_\nu(x)\,,
\end{equation}
with Hermite polynomials $H_\nu(x)$. In momentum space only one or two
states have a non-negligible eigenvalue $w_{\tilde\nu}$, therefore
higher optimized states, which do not contribute to the ground state 
of ${\cal H}$, are not expected to be converged. In all cases the 
most important states ($\tilde\nu=0$) resemble the eigenstates of
shifted Harmonic oscillators. If we use a single local set, the 
states correspond to symmetric and anti-symmetric combinations
of left- and right-shifted oscillator functions (remember the step
structure in Figure~\ref{figevals}). For $\omega_0 = 0.1\,t$ the
shift of the real space functions decreases with higher $\tilde\nu$.
This reflects the fact, that the lattice is slow and fermions move
within a distortion. In comparison, the case of high phonon 
frequency $\omega_0 = 10\,t$ is much simpler, as we deal only 
with states of almost fixed shift.

In summary, the simple example of the two-site Holstein model
provides good insight into the properties of the optimized phonon
approach. It is very efficient for determining an optimal basis
within a given phonon Hilbert space. Nevertheless, we are not relieved
from choosing the most appropriate decomposition of our model
under consideration. Here physical intuition and some knowledge
of the model is necessary to decide between real and momentum space,
or other special choices for the phonon Hilbert space. As demonstrated
above, the structure of the optimized wave functions and the 
behaviour of the eigenvalues of the density matrix may give some
hints. 
\section{Bipolaron band formation in the EHHM}
Besides bi-/polaron formation itself, the question of whether 
polarons or bipolarons can move itinerantly has been the subject 
of much controversy over the last decades (see, e.g., 
the debate in Ref.~\cite{BEMB92}). The existence of polaronic bands has 
been verified by ED techniques in 1D and 2D~\cite{St96,WRF96,WF97}, 
however, since both the width and the (electronic) spectral weight of
the polaronic bands are exponentially reduced in the strong-coupling 
case~\cite{FLW97}, the coherent band motion of small bi-/polarons 
becomes rapidly destroyed, e.g. by impurities or thermal fluctuations.

Recently it has been discovered that a longer-range EP interaction leads
to a decrease in the effective mass of polarons~\cite{AK99,FLW00}
and bipolarons~\cite{BT01} in the strong-coupling regime, which can
have significant consequences because the quasi-particles are more 
likely to remain mobile. Indeed, for the single-electron 
extended Holstein model~(\ref{ED1}) the polaron band dispersion 
was shown to be less renormalized as compared to the Holstein 
model~\cite{FLW00}. Here we would like to present some
results for the two-electron case, where the competition between
attractive EP interaction and Coulomb repulsion becomes important.   

\begin{figure}[!t]
  \begin{center}
    \epsfig{file= 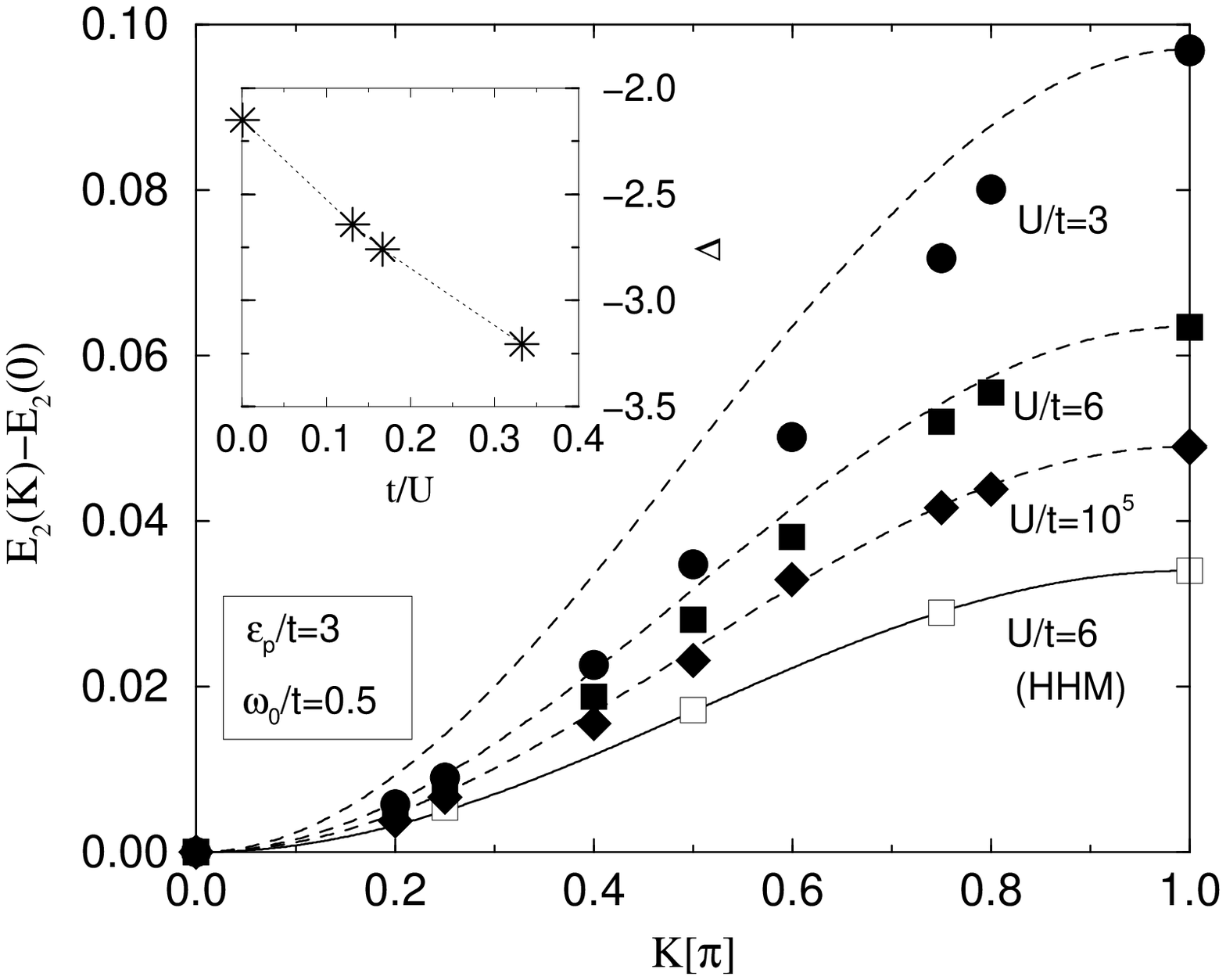, width=\linewidth}
  \end{center}
  \caption{
Bipolaron band structure in the 1D extended Holstein Hubbard model. 
Data points (filled symbols) are obtained from exact diagonalizations 
of the EHHM on eight- and ten-site chains (employing 
periodic boundary conditions) with different Hubbard interaction strengths; 
dotted lines give the corresponding rescaled cosine bands 
having the same bandwidths. At $U/t=6$, the bipolaron band 
dispersion of the Holstein Hubbard model is included
for comparison (open symbols). The inset displays the bipolaron binding 
energy $\Delta$ as a function of the inverse Hubbard interaction 
strength.}\label{figdis}
\end{figure}

By applying the optimized phonon approach outlined in Sec.~2 to 
EHHM we obtain the lowest eigenvalues of the 
two-particle system in each $K$ sector ($E_2(K)$), 
presented in Fig.~\ref{figdis}. The gain in  performance
compared to ordinary ED is illustrated in Table~\ref{tab1}. 
\begin{table}[b!]
\begin{minipage}{\textwidth}
\begin{tabular}[t]{|c|c|c|c|c|c|}\hline
&phonon&matrix &Lanczos& memory&CPU time \\ 
& cut-off & dimension&  diaganilizations    & requirement& per run  \\ \hline\hline
&  & &      & &    \\ 
\rb{ED} &  \rb{$\;\;M=81\;\;$}      &  \rb{$\sim 5 \times 10^{13}$} & \rb{1} & \rb{ $\gg 1000$ TBytes} & \rb{???}  \\ \hline\hline
optimized &  $M=13$  &   & &  & $\sim 20-200$  \\ 
phonons   &   $m=7$ &  \rb{$1.8 \times 10^{6}$} & \rb{50-300} & \rb{$\sim 500 $ MBytes} & CRAY T3E hrs  \\ \hline
\end{tabular}
\parbox[t]{.5cm}{\vspace{.5cm}}
\caption{ Problem sizes and computer requirements for the exact diagonalization method (ED) and the optimized phonon approach to solve one parameter set of the bipolaron problem on a ten-site lattice within the same accurancy. For both strategies the calculation of the ground-state in a fixed K-sector is assumed.}
\label{tab1}
\end{minipage}
\end{table}
The EP parameters  $\varepsilon_p/t=3.0$ and $\omega_0/t=0.5$ are chosen 
in order to address the intermediate coupling and frequency regime, 
which is almost impossible to investigate analytically.   
At first we note that the two electrons always form a bipolaronic bound state 
in the EHHM: The bipolaron binding energy $\Delta= E_2(0)-2 E_1(0)$ 
is negative, {\it irrespective} of the Hubbard interaction 
strength $U$ (see inset of Fig.~\ref{figdis}). 
This is an important difference between the EHHM and the HHM. 
In the HHM, a critical interaction $U_c$ exists for any EP coupling 
where the bipolaron unbinds~\cite{BT01,BKT00,WFWB00}.
Then, calculating the $K$-dependent binding energy, defined as  
$\Delta(K)=E_2(K)-\min_{K',K''}[E_1(K')+E_1(K'')]$ with 
\mbox{$K'+K''= K(\!\!\!\mod 2\pi)$}, we find a bound bipolaron  
for {\it all} $\;K$-values. This means that the dispersion curves 
depicted in Fig.~\ref{figdis} can be deemed to be well-defined 
bipolaron quasiparticle bands. Of course, due to the ``dressing'' with phonons,
the effective mass of the bipolaron is substantially enhanced.
Accordingly the coherent bandwidth of the EHHM bipolaron, 
$\Delta E=E_2(\pi)-E_2(0)$, is reduced as compared to the 
free electron case but, on the other hand, it is notably larger 
than those of the HHM bipolaron, indicating that the EHHM bipolaron 
is a rather mobile quasiparticle. As $U$ increases $\Delta E$
monotonously decreases, which clearly is a correlation effect
(put in mind that $U$ hinders double occupancy). 
At last we notice that the band structure
of the bipolaron significantly deviates from a rescaled cosine 
band only for $U$ much smaller than $2\tilde{\varepsilon}_p$. 
In this case the on-site and nearest-neighbour density-density correlation 
functions are about the same size. For $U \gtrsim 2\tilde{\varepsilon}_p$
the band becomes almost cosine shaped. This striking  cosine band dispersion,  
indicating free-particle like behaviour, was previously observed 
for the HHM model near $U = 2\varepsilon_p$ 
(Refs.~\cite{WRF96,WFWB00}, cf. also open symbols in Fig.~\ref{figdis}), 
and has been attributed to the formation of an {\it inter-site bipolaron}. 
If the inter-site bipolaron moves as a bound pair through the lattice, 
owing to the retardation effect ($\omega_0<t$),  
the second electron can take the advantage of the lattice distortion left 
by the first one, still avoiding the on-site Coulomb repulsion.  
As a result the residual bipolaron-phonon interaction is small. 
\begin{figure}[!b]
  \begin{center}
    \epsfig{file= 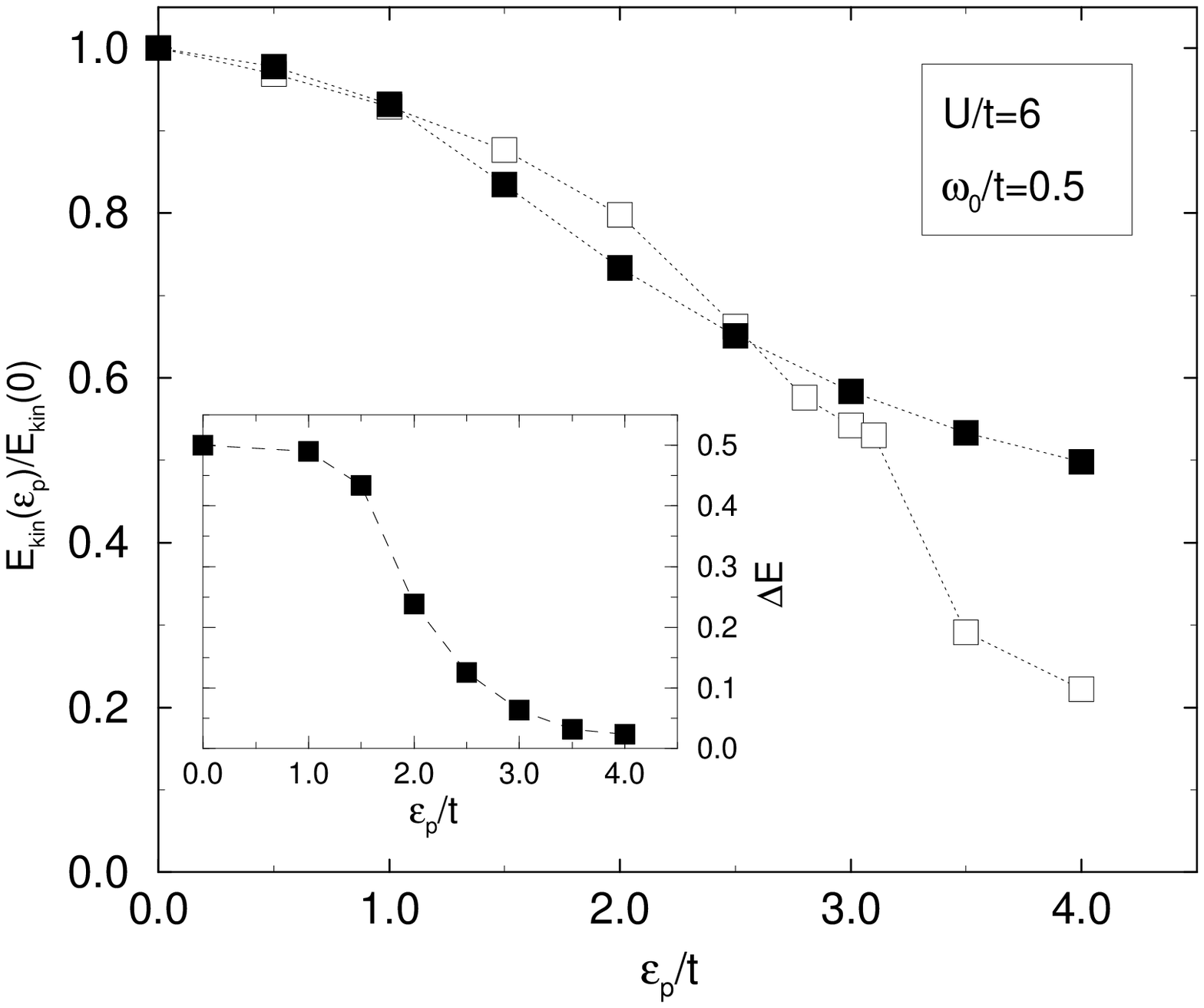, width=\linewidth}
  \end{center}
  \caption{Renormalized bipolaron kinetic energy in the eight-site  
EHHM (filled symbols) and HHM (open symbols). The inset gives the coherent
bandwidth of the bipolaron in dependence on the EP coupling 
strength.}\label{figekin}
\end{figure}

To gain more insight into the differences between EHHM and HHM 
bipolaronic states, we have calculated the bipolaron kinetic energy,
$E_{\rm kin}$, given by the ground-state average of the 
first term of~(\ref{ED1}).  
Figure~\ref{figekin} presents $E_{\rm kin}$ as a function of the EP 
interaction parameter $\varepsilon_p$ at fixed $U/t=6$ 
and $\omega_0/t=0.5$. Comparing the results for the EHHM and HHM, 
we can distinguish three regimes.  For small EP couplings,
the two polarons are bound in the EHHM, but not in the HHM.  
The higher kinetic energy of the EHHM bipolaron is mainly due to   
the larger incoherent contribution to the  f-sum rule, which originates
from incoherent hopping processes of the two charge carriers 
within the joint lattice distortion spread over the whole lattice. 
In the intermediate coupling regime predominantly inter-site 
bipolarons are formed in both models. Now the coherent part to 
the f-sum rule being proportional to the (inverse) effective mass)  
plays a decisive role. The EHHM bipolaron has to drag a larger phonon 
cloud than the HHM bipolaron coherently through the lattice 
and therefore acquires a 
larger effective mass. At strong EP couplings, in the 
HHM, a second transition to an on-site bipolaronic state takes place 
at about $\varepsilon_p/t=3 (\simeq U/2)$, whereas 
the EHHM bipolaron stays in a spatially much more extended state.
Consequently we observe a very gradual decrease of the bipolaron 
kinetic energy for the EHHM. Finally we comment on the renormalization 
of the coherent bandwidth of the EHHM bipolaron. As in the single polaron
case, at small EP couplings the band structure is flattened near the 
zone boundary by the intersection with the dispersionsless phonon branch. 
Hence we find  $\Delta E\simeq 0.5$ for $\varepsilon_p/t\ll 1$.
The strong reduction of the bandwidth starting to come in at about
$\varepsilon_p/t\gtrsim 1$ can be traced back to the formation
of a bipolaron with pronounced nearest-neighbour correlations.
\section{Summary}
The objective of this work was the presentation of an advanced phonon 
optimization algorithm for application in Lanczos diagonalizations.
At the heart of our procedure an 'optimized' phonon basis 
is created automatically, improving the most important eigenstates of the 
density matrix by means of gradual 'sweeps' through a sufficiently 
large part of the infinite dimensional phonon Hilbert space.
Within this scheme the density matrix is calculated in compliance with 
the symmetries of the underlying model.  
Depending on the physical problem under consideration, the efficiency 
of the proposed method might be improved considerably using more than one 
set of optimized phonon states or by performing the optimization 
procedure in momentum space.     

The reliability of our approach was demonstrated by calculating 
the band dispersion of bipolarons in the framework of the extended 
Holstein Hubbard model. In comparison with the Holstein Hubbard model 
where with increasing strength of the EP interaction  a sequence 
of transitions from two unbound large polarons to an inter-site 
bipolaron and finally to a self-trapped on-site bipolaron takes place, 
in the EHHM the two electrons form a bipolaronic bound state for all 
EP couplings, 
irrespective of the magnitude of the Hubbard interaction. 
As an effect of the longer ranged non-screened EP coupling 
included in the EHHM, the EHHM bipolaron is a rather 
mobile spatially extended quasiparticle even in the extreme 
strong-coupling regime. 

\bibliography{ref}
\bibliographystyle{phys}
\end{document}